\documentclass[aps,prb,twocolumn]{revtex4} 

\usepackage{graphicx}
\usepackage{dcolumn}
\usepackage{bm}
\usepackage{amsmath} 

\newcommand{\comment}[1]{}

\begin{document}

\title{Quantum Monte Carlo in Classical Phase Space
with the Wigner-Kirkwood Commutation Function.
Results for the Saturation Liquid  Density of $^4$He.}


\author{Phil Attard}
\affiliation{ {\tt phil.attard1@gmail.com}  1 Dec.\ 2025, \today }


\begin{abstract}
A Metropolis Monte Carlo algorithm is given
for the case of a complex phase space weight,
which applies generally in quantum statistical mechanics.
Computer simulations using Lennard-Jones $^4$He near the $\lambda$-transition,
including an expansion to third order
of the Wigner-Kirkwood commutation function,
give a saturation liquid density in agreement with measured values.
\end{abstract}


\maketitle

%
\section{Introduction}
\setcounter{equation}{0} \setcounter{subsubsection}{0}
\renewcommand{\theequation}{\arabic{section}.\arabic{equation}}
%

The computer simulation of quantum condensed matter
is a challenging problem.
A review of methods then extant has been given by Ceperley (1995).
His own quantum Monte Carlo algorithm
that implements the Feynman path integral over imaginary time
(equivalently, temperature slices) is the most impressive.
Ceperley (1995) illustrates his algorithm with extensive results
for saturated liquid $^4$He in the vicinity of the $\lambda$-transition.

One significant development since 1995
is the exact formulation of quantum statistical mechanics
in classical phase space (Attard 2016b, 2018b, 2021).
This has been used for Monte Carlo simulations of the noble gases
(Attard 2017),
with more recent focus on liquid helium-4
in the vicinity of the $\lambda$-transition (Attard 2025a, 2025b, 2025d).
Arguably,  Ceperley's (1995)
 Feynman path integral algorithm does a better job at handling
the non-commutativity of the position and momentum operators
(also known as the Wigner-Kirkwood commutation function,
also known as the Heisenberg uncertainty relation),
whereas Attard's classical phase space algorithm
is better at handling wave function symmetrization.

The computational burdens of the two approaches are very different.
Ceperley (1995) simulated 64 $^4$He atoms using a 1990's supercomputer.
Attard (2025a, 2025b, 2025d) simulated 5,000 $^4$He atoms
using a naughty personal computer.
The primitive approximation for the action on the Feynman path
requires about 1,000 temperature slices
(Ceperley 1995),
the cost of each one of which is comparable to a single classical simulation.
Ceperley (1995) gives more sophisticated approximations for the action
that require far fewer slices, ${\cal O}(20)$,
but the additional computational cost of each slice is unclear.

The purpose of the present paper is to improve
the classical phase space algorithm
with respect to the Wigner-Kirkwood commutation function,
whilst preserving its advantages
for the treatment of  wave function symmetrization,
and without increasing the computational burden.
A general Metropolis Monte Carlo algorithm
is given for complex phase space weights.
This is implemented for Lennard-Jones $^4$He on the saturation curve.
A fourth order temperature expansion
of the Wigner-Kirkwood commutation function is given,
and numerical results are obtained up to the third order.
Methods for including the symmetrization function are given,
but these are not implemented in the present paper.

%
\section{Formalism and Algorithm}
\setcounter{equation}{0} \setcounter{subsubsection}{0}
\renewcommand{\theequation}{\arabic{section}.\arabic{equation}}
%

\subsection{Phase Space Weight}

For a subsystem of  $N$ identical bosons in a volume $V$,
a point in classical phase space is
${\bf \Gamma} = \{{\bf q},{\bf p}\}$,
and its conjugate with reversed momenta is
${\bf \Gamma}^\dag = \{{\bf q},-{\bf p}\}$.
Here the position configuration is
${\bf q} = \{ {\bf q}_1,{\bf q}_2,\ldots, {\bf q}_N\}$
and the momentum configuration is
${\bf p} = \{ {\bf p}_1,{\bf p}_2,\ldots, {\bf p}_N\}$,
where the position of boson $j$ is
${\bf q}_j = \{ {q}_{jx},{q}_{jy},{q}_{jz}\}$,
and its momentum is
${\bf p}_j = \{ {p}_{jx},{p}_{jy},{p}_{jz}\}$.
Here we take the momentum to belong to the continuum;
see Attard (2025a, 2025b) for the treatment of quantized momentum.

For a canonical equilibrium system of temperature $T$ and volume $V$,
the phase space probability density is
(Attard 2018b, 2021)
\begin{equation}
\wp({\bf \Gamma}) =
\frac{
e^{-\beta {\cal H}({\bf \Gamma})}
e^{W({\bf \Gamma})} \eta({\bf \Gamma})
 }{ N! h^{3N}Z },
\end{equation}
where $\beta = 1/k_{\rm B}T$,
$k_{\rm B}$ 
being Boltzmann's constant,
$h$ 
is Planck's constant,
and the partition function, $Z(N,V,T)$,
which normalizes the probability,
gives the total entropy.
Also the classical Hamiltonian is
${\cal H}({\bf \Gamma}) = {\cal K}({\bf p}) + U({\bf q})$,
where ${\cal K}({\bf p}) = p^2/2m = \sum_{j=1}^N p_j^2/2m$
is the kinetic energy,
$m$ being the mass of a boson.
Below we shall take the potential energy
to consist of central pair potentials
$U({\bf q}) = \sum_{j<k}^N u(q_{jk})$.

\subsubsection{Commutation Function}

The main focus of this work is on the Wigner-Kirkwood
commutation function (Attard 2018b, 2021, Kirkwood 1933, Wigner 1932),
which is defined by
\begin{equation} \label{Eq:Wdefn}
e^{-\beta {\cal H}({\bf \Gamma})}
e^{W({\bf \Gamma})}
=
e^{{\bf p}\cdot{\bf q}/{\rm i} \hbar}
e^{-\beta \hat{\cal H}({\bf q})}
e^{-{\bf p}\cdot{\bf q}/{\rm i} \hbar} .
\end{equation}
This corresponds to $\omega_p =e^{W_p}$ (Attard 2018b Eq.~(2.6))
(present notation).
For the quantum weight one can use either $\omega_p \eta_q$
or else $\omega_q \eta_p$,
with $\omega_q^* = \omega_p$ and $\eta_q^* = \eta_p$.

On the right hand side of the definition of the commutation function
the Hamiltonian operator appears,
$ \hat{\cal H}({\bf q}) =  \hat{\cal K}({\bf q}) +  U({\bf q})$.
The Fourier factors are unnormalized, unsymmetrized momentum eigenfunctions.
The commutator $[\hat{\cal K}, U] \ne 0$
makes $W \ne 0$,
which reflects the Heisenberg uncertainty relation.

The Wigner-Kirkwood commutation function is complex
with the property that $W({\bf \Gamma}^\dag) = W({\bf \Gamma})^*$.
Separating the real and imaginary parts,
$W = W_{\rm r} + {\rm i} W_{\rm i}$,
we have
\begin{equation}
e^{W({\bf \Gamma})} =
e^{W_{\rm r}({\bf \Gamma})}
[ \cos W_{\rm i}({\bf \Gamma}) + {\rm i} \sin W_{\rm i}({\bf \Gamma}) ] .
\end{equation}
The imaginary part of this is odd in momentum and averages to zero
because in an equilibrium system forward and backward momentum
must be equally likely,
$\wp({\bf \Gamma}) = \wp({\bf \Gamma}^\dag)$.
Thus we only require the real part
to obtain averages of real functions.
For the moment we neglect the fact that
the symmetrization function $\eta$ is complex.
The commutation function
becomes non-zero at higher temperatures
than the symmetrization function,
and so we can deduce intrinsic properties of the former
in the absence of the latter.

The cosine term rapidly fluctuates about zero,
particularly since $W$ is an extensive thermodynamic variable.
Such rapid oscillations cancel each other,
and the corresponding region of phase space has zero weight.
Therefore, we impose the restriction
\begin{equation} \label{Eq:Wi>pi/2}
|W_{\rm i}({\bf \Gamma})| < \frac{\pi}{2} .
\end{equation}
The justification for this is that
at high temperatures the system is classical
and $W({\bf \Gamma})=0$.
Presumably, as the temperature is lowered
the system never crosses the boundary where $\cos W_{\rm i}({\bf \Gamma})$
changes sign.
As will shortly be shown, the Wigner-Kirkwood commutation function
depends upon the gradients of the pair potential.
Since these become large in the repulsive core region of the pair potential,
this restriction prevents particles approaching each other too closely.
(Some gradients appear as the scalar product with the momentum,
and so this also tends to lower the magnitude of the latter.)
As the temperature is lowered, higher order gradients come into play,
and these are large at greater separations than lower order gradients.
This restriction is a manifestation of the Heisenberg uncertainty relation:
it keeps particles further apart than classical considerations alone imply.
This condition is perhaps best judged by the results that follow.

A standard approach to the simulation of classical equilibrium systems
is Metropolis Monte Carlo (Allen  and Tildesley  1987).
In the most general form
this says that a trial move to a new configuration
should be accepted if $\wp^{\rm new}/\wp^{\rm old} \ge r$,
where $r$ is a random number uniformly distributed on $[0,1]$.
In the present case this says accept the trial configuration if
\begin{equation}
e^{-\beta \Delta {\cal H}({\bf \Gamma})}
e^{\Delta W_{\rm r}({\bf \Gamma})}
\frac{\cos W_{\rm i}({\bf \Gamma}^{\rm new})
}{
\cos W_{\rm i}({\bf \Gamma}^{\rm old}) }
\ge r.
\end{equation}
This is the major result of this paper.
Note that both position and momentum are changed in a trial move,
which is generally done one particle at a time.
In addition, the trial configuration is rejected
if the restriction  (\ref{Eq:Wi>pi/2}) is violated.
Hence if we start with $\wp > 0$,
then this means that the probability always remains positive.

Starting with an initial phase space point ${\bf \Gamma}_0$,
we randomly change the signs of the ${\bf p}_j$
until we find a point that satisfies the restriction (\ref{Eq:Wi>pi/2}).
A similar procedure is performed
if the starting configuration is taken from one equilibrated
at a different temperature or density.
Usually this takes on the order of $10^2$--$10^3$ random changes
for $N=$1--5$\times 10^3$ atoms.
One should then equilibrate the subsystem before collecting statistics.

\subsubsection{Symmetrization Function}

In the computational results for $^4$He presented below
the symmetrization function is neglected.
This is appropriate on the high temperature side of the $\lambda$-transition.
Our main aim is to delineate the r\^ole of the commutation function,
and we shall take up the combined effects of the two in future work.
For completeness, here we show one way in which this can be done.

The symmetrization function
is the ratio of unpermuted to permuted momentum eigenfunctions
summed over all permutations  (Attard 2018b, 2021),
\begin{equation}
\eta({\bf \Gamma})
=
\sum_{\hat{\rm P}}
e^{-[{\bf p}-\hat{\rm P}{\bf p}]\cdot{\bf q}/{\rm i}\hbar} .
\end{equation}
This corresponds to $\eta_q$  (Attard 2018b Eq.~(2.4)).
That part of the grand potential due to symmetrization
is given by the average of this,
\begin{eqnarray}
e^{-\beta \Omega_{\rm sym}}
& = &
\langle \eta({\bf \Gamma}) \rangle_W
\nonumber \\ & = &
e^{ \langle \stackrel{\circ}{\eta}({\bf \Gamma}) \rangle_W } .
\end{eqnarray}
In the second equality,
which is believed to be exact in the thermodynamic limit
(Attard 2018b \S~III\,B\,1),
the sum over single permutation loops is
$\stackrel{\circ}{\eta}\!\!({\bf \Gamma}) =
\sum_{l=2}^\infty \eta^{(l)}({\bf \Gamma})$,
where the $l$-loop symmetrization function is
\begin{equation}
\eta^{(l)}({\bf \Gamma})
=
\sum_{j_1,\ldots,j_l}^N\!\!\!'\; \prod_{k=1}^{l}
e^{ - {\bf p}_{j_k} \cdot {\bf q}_{j_k,j_{k+1}} /{\rm i}\hbar}
 , \quad j_{l+1} \equiv j_1 .
\end{equation}
The sum is over the unique directed cyclic permutations
of all subsets of $l$-bosons.

The $l$-loop symmetrization function is complex,
$\eta^{(l)}({\bf \Gamma})
=
\eta^{(l)}_{\rm r}({\bf \Gamma})
+ {\rm i} \eta^{(l)}_{\rm i}({\bf \Gamma})$.
When averaged it gives the $l$-loop grand potential.
Again because the imaginary part is odd in momentum,
only the real part of the integrand survives,
$\mbox{Re} \{e^{{\rm i}W_{\rm i}(\Gamma)} \eta^{(l)}(\Gamma)\}
=
\cos \big( W_{\rm i}(\Gamma)\big) \eta^{(l)}_{\rm r}({\bf \Gamma})
- \sin \big( W_{\rm i}(\Gamma)\big) \eta^{(l)}_{\rm i}({\bf \Gamma})$.
(This is why it is important to combine
$W_p$ with $\eta_q$, or else  $W_q$ with $\eta_p$.)
Hence the $l$-loop grand potential is
(Attard 2018b, 2021)
\begin{eqnarray}
-\beta \Omega_W^{(l)}
&=&
\langle \eta^{(l)}(\Gamma) \rangle_W
 \\ & = &
\left\langle
\big[ \eta^{(l)}_{\rm r}({\bf \Gamma})
- \tan \!\big( W_{\rm i}(\Gamma)\big) \eta^{(l)}_{\rm i}({\bf \Gamma}) \big]
\right\rangle_W
\nonumber \\ & = &\nonumber
\frac{
\int {\rm d}{\bf \Gamma} \,
e^{-\beta {\cal H}} e^{W_{\rm r}}
\big[ \eta^{(l)}_{\rm r} \cos  W_{\rm i}
- \eta^{(l)}_{\rm i} \sin  W_{\rm i} \big]
}{
\int {\rm d}{\bf \Gamma} \,
e^{-\beta {\cal H}} e^{W_{\rm r}} \cos  W_{\rm i}
}.
\end{eqnarray}
(The monomer grand potential is just the logarithm
of the partition function in the absence of the symmetrization function.)
Several tricks have been found to facilitate
the computation of the  $l$-loop symmetrization function
(Attard 2021 \S5.4.2).
These will undoubtedly prove useful in the present more complex case
of numerical momentum quadrature.
On the high temperature side of the $\lambda$-transition,
we expect that only position loops with consecutive particles
separated by less than about the thermal wavelength will contribute.

\subsection{Fluctuation Expansion for the Commutation Function}

Kirkwood (1933) took the inverse temperature derivative
of the defining equation for the commutation function, Eq.~(\ref{Eq:Wdefn}),
to obtain an infinite series in powers of $\beta$,
explicitly giving the first non-zero term, which is quadratic in $\beta$.
Higher order coefficients using this method have been obtained
(Attard 2021 \S\S8.2--8.4).

A slightly different approach based on a fluctuation expansion
has $n$th order coefficient (Attard 2021 \S8.5)
\begin{equation}
\Delta_{\cal H}^{(n)}({\bf q},{\bf p})
\equiv
\frac{ \langle {\bf q} | \big[ \hat{\cal H}
- {\cal H}({\bf q},{\bf p}) \big]^n | {\bf p} \rangle
}{ \langle {\bf q} |  {\bf p} \rangle }  .
\end{equation}
One has $\Delta^{(0)}_{\cal H}({\bf q},{\bf p}) = 1$
and $\Delta^{(1)}_{\cal H}({\bf q},{\bf p}) = 0$.
These give a series for the exponent of the form
\begin{equation}
W(\Gamma) =
\frac{\beta^2}{2!} \tilde \Delta_{\cal H}^{(2)}(\Gamma)
-\frac{\beta^3}{3!} \tilde \Delta_{\cal H}^{(3)}(\Gamma)
+ \frac{\beta^4}{4!} \tilde \Delta_{\cal H}^{(4)}(\Gamma)
-\ldots
\end{equation}
The second order fluctuation,
$\tilde \Delta^{(2)}_{\cal H} = \Delta^{(2)}_{\cal H}$, is
\begin{eqnarray}
\Delta^{(2)}_{\cal H}({\bf q},{\bf p})
 & = &
\frac{-\hbar^2}{2m} \nabla^2 U({\bf q})
 - \frac{\mathrm{i}\hbar}{m} {\bf p} \cdot \nabla  U({\bf q}) .
\end{eqnarray}
The third order fluctuation,
$\tilde \Delta^{(3)}_{\cal H} = \Delta^{(3)}_{\cal H}$, is
\begin{eqnarray}
\Delta^{(3)}_{\cal H}({\bf q},{\bf p})
& = &
\frac{- \hbar^2}{m}  \nabla U\! \cdot\!  \nabla U
+ \frac{\hbar^4}{4m^2} \nabla^2 \nabla^2 U
 \\ && \mbox{ }\nonumber
+ \frac{\mathrm{i}\hbar^3}{m^2} {\bf p} \!\cdot\! \nabla \nabla^2 U
- \frac{\hbar^2}{m^2} {\bf p} {\bf p}  : \nabla \nabla  U .
\end{eqnarray}
And the fourth order fluctuation  is
\begin{eqnarray}
\tilde \Delta^{(4)}_{\cal H}({\bf q},{\bf p})
& = &
\frac{5\hbar^4}{2m^2}  \nabla U \cdot \nabla \nabla^2 U
+ \frac{5\mathrm{i}\hbar^3}{m^2} {\bf p}\nabla U : \nabla \nabla U
\nonumber \\ && \mbox{ }
+ \frac{\hbar^4}{m^2}  \nabla \nabla  U \!:\! \nabla \nabla U
- \frac{\hbar^6}{8m^3} \nabla^2 \nabla^2  \nabla^2 U
\nonumber \\ && \mbox{ }
- \frac{3\mathrm{i}\hbar^5}{4m^3} {\bf p}\!\cdot\!\nabla \nabla^2 \nabla^2 U
+ \frac{3\hbar^4}{2m^3} {\bf p}{\bf p}: \nabla\nabla \nabla^2 U
\nonumber \\ && \mbox{ }
+ \frac{\mathrm{i}\hbar^3}{m^3}  {\bf p}{\bf p}{\bf p}
\vdots \nabla\nabla\nabla U .
\end{eqnarray}
Expressions for the gradients of a central pair potential
have been catalogued
(Attard 2021 \S\S 9.5.2 and 9.5.3).

The order at which the expansion is terminated is denoted $n_W^{\rm max}$.
The value  $n_W^{\rm max}=0$ corresponds to a classical simulation.

\subsection{Temperature Derivatives}

The inverse temperature derivative
of the logarithm of the partition function,
without $\eta$, is
\begin{eqnarray}
\tilde E
& \equiv &
\frac{ -\partial \ln Z}{\partial \beta}
\nonumber \\ & = &
\frac{-1}{2 Z}
\int {\rm d}{\bf \Gamma}\, \left\{
e^{-\beta {\cal H}} 
e^{ W_{\rm r}}
e^{ {\rm i}W_{\rm i}}
\left[  \dot W_{\rm r} - {\cal H} + {\rm i} \dot W_{\rm i} \right]
\right. \nonumber \\ && \left. \mbox{ }
+
e^{-\beta {\cal H}}
e^{ W_{\rm r}}
e^{-{\rm i}W_{\rm i}}
 \left[  \dot W_{\rm r} - {\cal H} - {\rm i}\dot W_{\rm i}
 \right] \right\}
\nonumber \\ & = &
\frac{-1}{ Z}
\int {\rm d}{\bf \Gamma}\,
e^{-\beta {\cal H}}
e^{ W_{\rm r}}
\left\{
[ \dot W_{\rm r} -{\cal H} ] \cos   W_{\rm i}
-\dot W_{\rm i} \sin W_{\rm i}
\right\}
\nonumber \\ & = &
\left\langle {\cal H}({\bf \Gamma})
- \dot W_{\rm r}({\bf \Gamma})
+ \dot W_{\rm i}({\bf \Gamma}) \tan W_{\rm i}({\bf \Gamma})
\right\rangle_W .
\end{eqnarray}
The over-dot signifies the inverse temperature derivative,
$\partial/\partial\beta$.
The second derivative is
\begin{eqnarray}
\frac{ \partial \tilde E}{\partial \beta}
& = &
\frac{(\partial Z/\partial\beta)^2}{Z^2}
- \frac{1}{2Z}
\int {\rm d}{\bf \Gamma}\,
e^{-\beta {\cal H}({\bf \Gamma})}
e^{ W_{\rm r}({\bf \Gamma})}
e^{ {\rm i}W_{\rm i}({\bf \Gamma})}
\nonumber \\ && \mbox{ } \times
\left\{[-{\cal H} + \dot W_{\rm r}+{\rm i} \dot W_{\rm i}]^2
+ [\ddot W_{\rm r}+{\rm i} \ddot W_{\rm i} ] \right\}
+{\rm cc}
\nonumber \\ & = &
\left\langle {\cal H} - \dot W_{\rm r}
+  \dot W_{\rm i}\tan  W_{\rm i} \right\rangle^2_W
\nonumber \\ && \mbox{ }
-
\left\langle {\cal H}^2 + \dot W_{\rm r}^2 - \dot W_{\rm i}^2
- 2 {\cal H} \dot W_{\rm r} + \ddot W_{\rm r}
\right. \nonumber \\ && \left.\mbox{ }
-  [-2 {\cal H} \dot W_{\rm i}
+ 2 \dot W_{\rm r}\dot W_{\rm i}
+  \ddot W_{\rm i}]\tan  W_{\rm i}
\right. \nonumber \\ && \left.\mbox{ }
 - \dot W_{\rm i}^2 \sec^2  W_{\rm i} 
\right\rangle_W .
\end{eqnarray}
(The final term was missing in version 1.)
Note that the heat capacity is $C_V = \partial \tilde E/\partial T
= -k_\mathrm{B} \beta^2 \partial \tilde E/\partial \beta$.

%
\section{Computational Results}
\setcounter{equation}{0} \setcounter{subsubsection}{0}
\renewcommand{\theequation}{\arabic{section}.\arabic{equation}}
%

\subsection{Model and Simulation Details}

The Lennard-Jones pair potential was used,
$u(r) = 4 \varepsilon[(\sigma/r)^{12} - (\sigma/r)^{6} ]$,
with helium parameters,
$ \varepsilon_{\rm He} = 10.22 k_{\rm B}$\,J and
$\sigma_{\rm He} =0.2556$\,nm (van Sciver 2012).
The potential was  cut-off at $R_{\rm cut} = 3.5\sigma$.
Quantum Monte Carlo simulations were performed in classical phase space
with either 1,000 or 5,000 atoms.
A low total density, $\rho \sigma^3 \approx $ 0.02--0.2,
allowed a droplet of density $\rho \sigma^3 \approx $ 0.3--0.9
to condense in the center of the system surrounded by its vapor.
Periodic boundary conditions were applied.
A small cell spatially based neighbor table was used.
The position and momentum steps for a trial configuration
were chosen to give acceptance rates of 30--60\%.

As the temperature is decreased,
the higher order gradients of the potential
in the expansion for the commutation function come into play.
These become very large at smaller separations.
It was found that occasionally pairs of particles
 became trapped in the core region,
presumably due to numerical overflow.
The problem more commonly arose when changing density or temperature,
and using the previous equilibrated configuration to start.
The effect could be seen in the form
of noise in the otherwise zero core region of the
radial distribution function;
this had amplitude ${\cal O}(10^{-3})$,
compared to its peak height of ${\cal O}(10^{0})$.
Although this has negligible effect on the statistics,
there is no need to calculate zero explicitly.
Therefore a hard core was implemented,
with value $q_{\rm min} =$ 1.1--1.4, depending on the temperature.
Any trial move that bought an atom closer to a neighbor than this
was rejected.
It was checked at the end of each simulation
that the radial distribution went smoothly to zero
and that the first peak was well-separated from the imposed hard-core.

For the fourth order case, $n_W^{\rm max}=4$,
the commutation function became very large,
${\cal O}(10^{4})$,
when any two particles came close together, $q_{jk} \alt 1.6\sigma$.
This was due to the terms with
a product of the gradients of the pair potential.
Unfortunately the change in the exponent of the probability ratio
was similarly large and led to numerical overflow.
The problem could probably have been overcome
by using a larger minimum separation $q_{\rm min}$,
or a smaller position step to obtain the trial position configuration.
However the former was deemed uncomfortably large,
and the latter would have meant an unacceptably lengthy simulation.
In so far as the fourth order term is a fluctuation of a fluctuation,
(Attard 2021 Eq.~(8.98)),
it and higher order terms may well be negligible.
No results for $n_W^{\rm max}=4$ are reported below.


\subsection{Results}

\begin{table}[tb]
\caption{
Simulation results for
the density $\rho \sigma^3$ and kinetic energy $\beta {\cal K}/N$
of liquid $^4$He on the saturation curve
using the Wigner-Kirkwood commutation function
with $ n_W^{\rm max}=$0, 2, or 3 ($N=1,000$).
\label{Tab:dens} }
\begin{center}
\begin{tabular}{c c c c c c c c c}
\hline\noalign{\smallskip}
$k_\mathrm{B} T /\varepsilon $ & $\Lambda/\sigma$ &
0$^a$  &  2  & 3  &
$\rho_{\rm meas}^{\rm sat} \sigma^3$ &
$ \beta {\cal K}/N$$^b$    \\
\hline 
0.8  & 1.1940 & 0.8023 & 0.30(1)  & 0.1912(6) &-& 1.1878(5) \\
0.7  & 1.2764 & 0.8470 & 0.398(2) &  0.2027(6) &-& 1.1505(6)\\
0.6  & 1.3787 & 0.8872 & 0.446(2) &  0.229(1) &-& 1.0930(4)\\
0.5  & 1.5103 & 0.9331 & 0.483(4) & 0.2743(5)& 0.255$^c$ & 0.975(1) \\
0.45 & 1.5920 & 0.96 &     -      & 0.3236(5) & 0.289 & 0.728(1)  \\
0.45$^d$ &   -    &    - &   -    & 0.3216(3) &   -   & 0.705(1)  \\
0.4  & 1.6886 & 0.98 &     -      & 0.34(2) & 0.317 & 0.642(5)   \\
0.35 & 1.8051 &   -  &     -      & 0.32(1) & 0.339 & 0.580(2)\\
0.3  & 1.9498 &   -  &    -       & 0.32(7) & 0.355 & 0.501(1)\\
0.25 & 2.1359 &  -   &   -        & 0.3324(1) & 0.364 & 0.454(3)\\
\hline
\end{tabular}
\end{center}
\flushleft
$^a$Classical simulations
(Attard 2022a, 2025a).  
$^b$$ n_W^{\rm max}=3$.
$^c$Extrapolated,
$k_\mathrm{B} T^{\rm crit}_{\rm meas} /\varepsilon = 0.48$.
$^d$$N=5,000$.
\end{table}

Simulation results are shown in Table~\ref{Tab:dens}.
For the classical simulation, $ n_W^{\rm max}=0$,
the saturated liquid density is about three times the measured value.
Including the commutation function at second order
reduces the density by about a factor of two,
and including it to third order reduces it to about the measured value.

The kinetic energy per particle is $3k_{\rm B}T/2$ in the classical case
and also for $ n_W^{\rm max}=  2$.
In the third order case the kinetic energy is reduced from this value,
increasingly so as the temperature is decreased.

A number of simulations were carried out with $N=5,000$,
and the results of one such is shown in the table.
No size dependence was observed.
Similarly, simulations were repeated with different over-all densities
without observing any systematic effects.

The measured critical temperature of $^4$He
in Lennard-Jones units is
$k_\mathrm{B} T^{\rm crit}_{\rm meas} /\varepsilon = 0.48$.
Table~\ref{Tab:dens}
gives results for the liquid saturation density above this temperature.
In these cases a liquid droplet in a gas phase
could be discriminated in the system.
Evidently the critical temperature for the Lennard-Jones model
is higher than the measured critical temperature,
and  it depends on the order of the commutation function expansion.

For $k_{\rm B}T/\varepsilon=0.50$ and  $\rho_{\rm tot}\sigma^3=0.26$
(droplet, central density 0.290(9), edge oscillations),
gives $\beta \overline E/N = -12.41(4)$
and $-\beta^2 \partial \overline E/N\partial \beta = 34(2)$.
For $k_{\rm B}T/\varepsilon=0.49$,
$\beta \overline E/N = -13.41(4)$
and $-\beta^2 \partial \overline E/N\partial \beta= 37(1)$.
(These results have been recalculated from version~1.)

The numerical difference is
$\Delta E /Nk_{\rm B}\Delta T = 37.06(3)$.
That this agrees with the directly simulated heat capacity
is a very sensitive test of the mathematical expressions
for the energy and the heat capacity,
and of their computer implementation.
It also tests the Metropolis Monte Carlo algorithm
and its implementation.
The author can confirm 
that it takes many rounds of debugging
to get quantitative agreement between the two.

The energy and heat capacity are sensitive to the
saturated droplet and  vary with the total density.
For example, at $k_{\rm B}T/\varepsilon=0.50$,
an overall density $\rho_{\rm tot}\sigma^3=0.15$.
forms a smooth droplet with central density 0.28
and  gives $\beta \overline E/N = -10.37(1)$.
An overall density $\rho_{\rm tot}\sigma^3=0.03$
forms a smooth droplet with central density 0.28
and gives $\beta \overline E/N = -8.6(1)$.
In these the energy of all atoms in the system are included.

\begin{figure}[t]
\centerline{ \resizebox{8cm}{!}{ \includegraphics*{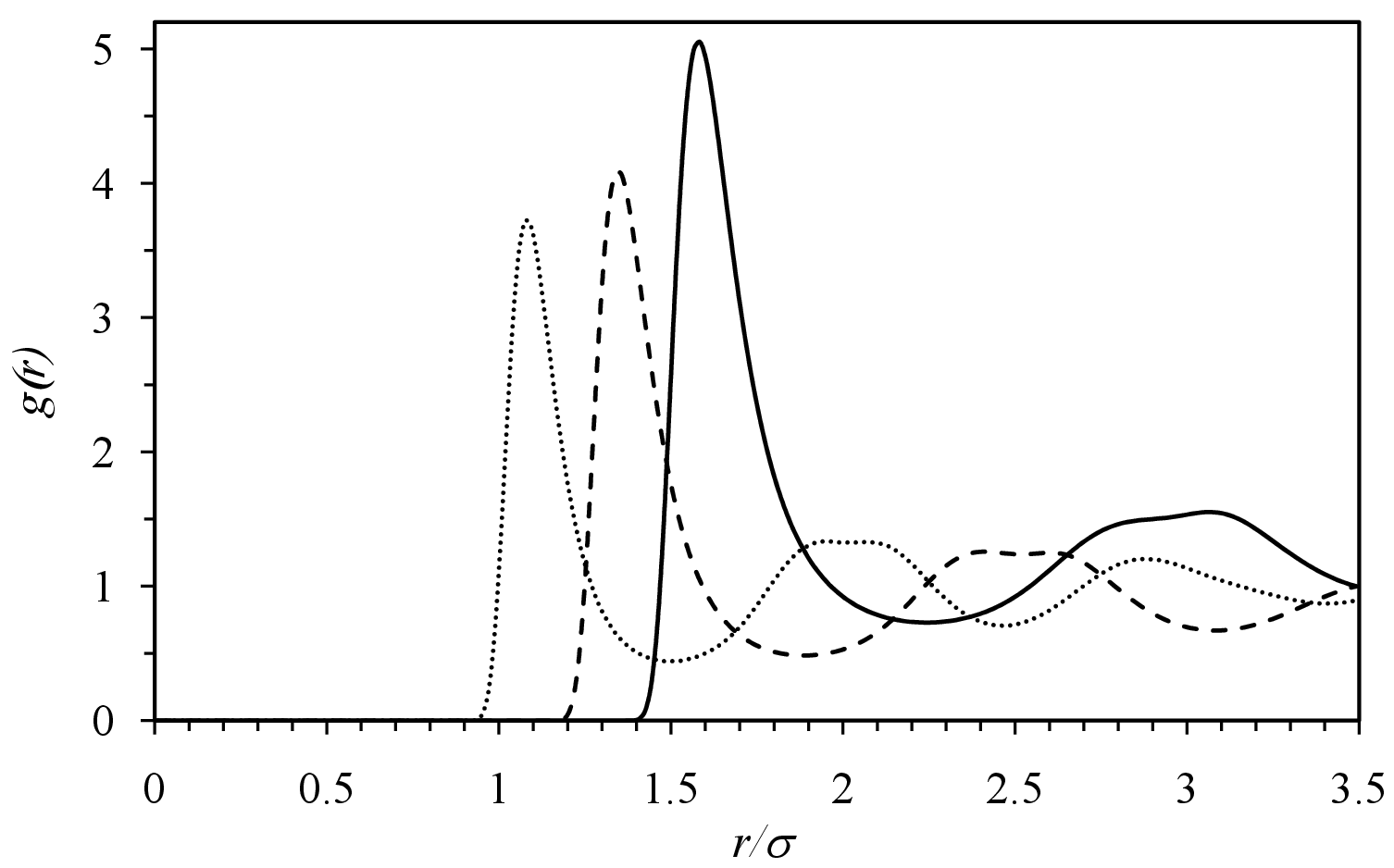} } }
\caption{\label{Fig:g(r)}
Radial distribution function for Lennard-Jones $^4$He
at $k_{\rm B}T/\varepsilon =0.5$
at the respective liquid saturation densities.
The dotted curve is $n_W^{\rm max} = 0$
(classical, $\rho\sigma^3=0.9331$),
the dashed curve is $n_W^{\rm max} = 2$
($\rho\sigma^3=0.483(4)$, $q_{\rm min} = 1.1\sigma$),
and the dotted curve is $n_W^{\rm max} = 3$
($\rho\sigma^3=0.2790(2)$, $q_{\rm min} = 1.3\sigma$).
}
\end{figure}

Figure~\ref{Fig:g(r)} shows the radial distribution function
for different orders of the commutation function.
The statistics were collected over the whole system,
but they are dominated by the saturated liquid phase.
It can be seen that in the second and third order cases,
the peak of the distribution occurs at a significantly greater
separation than the imposed hard-core diameter,
and that the distribution goes smoothly to zero.
The signature of using too large a value for the hard-core
is that the distribution goes discontinuously to zero
immediately from the peak.

Notice how the peak of the radial distribution function
shifts to larger separations
as the order of the expansion for the commutation function is increased.
This is a manifestation of the Heisenberg uncertainty relation,
which smears out or delocalises bound particles.
This is the origin of the reduction in the saturated liquid density
compared to the classical case.

\begin{figure}[t]
\centerline{ \resizebox{8cm}{!}{ \includegraphics*{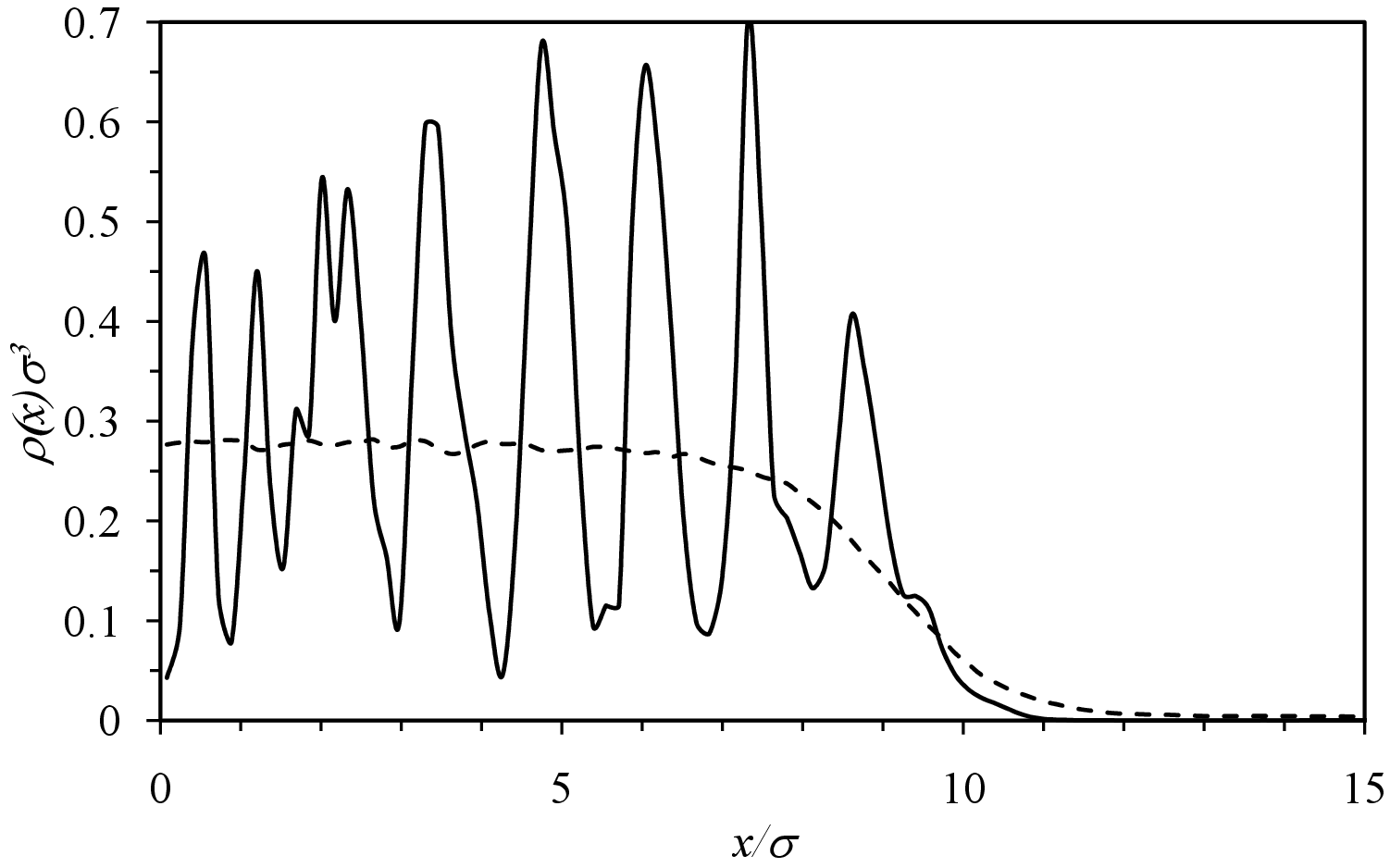} } }
\caption{\label{Fig:rho(x)}
Density profile along an axis through the system
at $k_{\rm B}T/\varepsilon =0.5$ (dashed curve)
and at $k_{\rm B}T/\varepsilon =0.45$ (solid curve)
for $n_W^{\rm max}=3$.
The statistical error (95\% confidence)
is about 4\% for the former,
and about 20\% for the latter.
}
\end{figure}

Figure~\ref{Fig:rho(x)}
shows the density in cross-section of the system at two temperatures.
The density was averaged in a cylinder of radius $\sigma$
about an axis.
It can be seen that at $k_{\rm B}T/\varepsilon =0.5$
the droplet is liquid-like,
and that at $k_{\rm B}T/\varepsilon =0.45$ it is solid-like.
In the latter case there is still substantial diffusion
of the atoms over the course of the simulation.
This and the fact that the troughs don't go to zero
means that it is not a rigidly crystalline state.
Of course, even a liquid displays density oscillations
at the liquid-vapor interface.
The reason for describing the state as solid-like
is the qualitative change in density profiles
over such a small temperature change.
As the temperature is further reduced,
the peaks become more pronounced,
the troughs can go to zero,
and the diffusion  becomes more limited.
The density quoted in Table~\ref{Tab:dens}
is the average over a sphere of radius $8\sigma$
about the origin.

Real $^4$He does not become solid
at saturation pressure down to absolute zero,
but it does solidify at elevated pressures.
One might speculate that the present solid-like state
results from a combination of three effects.
The first is the approximate nature
of the Lennard-Jones pair potential
together with the termination
of the temperature expansion for the commutation function
at $n_W^{\rm max}=3$.
The second is the Laplace pressure acting
on the interior of the nano-droplet.
The third is the periodic boundary conditions.

%
\section{Conclusion}
\setcounter{equation}{0} \setcounter{subsubsection}{0}
\renewcommand{\theequation}{\arabic{section}.\arabic{equation}}
%

A benefit of the present third order approximation
for the Wigner-Kirkwood commutation function
is that allows the classical phase space simulation
of Lennard-Jones $^4$He
at about the measured saturation liquid density
in a homogeneous system. 
In the absence of the commutation function
such a system would cavitate
into a liquid drop at the bare Lennard-Jones density,
$\rho_{LJ}^{\rm sat} \sigma^3 \approx 0.9$, and a vapor phase.
Being able to use the measured saturation liquid  density
for a homogeneous simulation is a significant advance.

\section*{References}


\begin{list}{}{\itemindent=-0.5cm \parsep=.5mm \itemsep=.5mm}

\item 
Allen M P and Tildesley D J 1987
\emph{Computer Simulation of Liquids}
(Oxford: Clarendon Press)

\item 
Attard P 2016b
Quantum statistical mechanics as an exact classical expansion with
results for Lennard-Jones helium
arXiv:1609.08178v3

\item
Attard P 2017
Quantum statistical mechanics results for argon, neon, and helium using
classical Monte Carlo
arXiv:1702.00096

\item 
Attard P 2018b
Quantum statistical mechanics in classical phase space. Expressions for
the multi-particle density, the average energy, and the virial pressure
arXiv:1811.00730

\item 
Attard P  2021
\emph{Quantum Statistical Mechanics in Classical Phase Space}
(Bristol: IOP Publishing)

\item 
Attard P 2025a
\emph{Understanding Bose-Einstein Condensation,
Superfluidity, and High Temperature Superconductivity}
(London: CRC Press)

\item 
Attard P 2025b
The molecular nature of superfluidity: Viscosity of helium from quantum
stochastic molecular dynamics simulations over real trajectories
arXiv:2409.19036v5

\item 
Attard P 2025d
Bose-Einstein condensation and the lambda transition
for interacting Lennard-Jones helium-4
arXiv:2504.07147v3

\item 
Ceperley  D M  1995
Path integrals in the theory of condensed helium
\emph{Rev.\ Mod.\ Phys.}\ {\bf 67} 279

\item 
Kirkwood J G 1933
Quantum statistics of almost classical particles
\emph{Phys.\ Rev.}\ {\bf 44}, 31

\item 
van Sciver  S W 2012
\emph{Helium Cryogenics}
(New York: Springer 2nd edition)

\item 
Wigner E 1932
On the quantum correction for thermodynamic equilibrium
\emph{Phys.\ Rev.}\ {\bf 40}, 749

\end{list}

\end{document}